\documentclass[twocolumn,pra,superscriptaddress,showpacs,amsmath,amssymb]{revtex4}

\usepackage{graphicx, amsmath, upgreek, amssymb}

\newcommand{\eref}[1]{Eq.~(\ref{#1})}

\newcommand{\fref}[1]{Fig.~\ref{#1}}
\newcommand{\Fref}[1]{Fig.~\ref{#1}}

\newcommand{\Sref}[1]{Section~\ref{#1}}

\newcommand{\force}{F}

\DeclareMathOperator{\Tr}{Tr}
\newcommand{\im}[1]{\,\text{Im}\!\left\{#1\right\}}
\newcommand{\re}[1]{\,\text{Re}\!\left\{#1\right\}}

\makeatletter
\newlength \figwidth
\makeatother

\begin{document}


\title[Scattering theory of multilevel atoms]{Scattering theory of multilevel atoms interacting with arbitrary radiation fields}

\author{Andr\'e Xuereb}
\affiliation{School of Physics and Astronomy, University of Southampton, Southampton SO17~1BJ, United Kingdom}
\email[To whom all correspondence should be addressed. Electronic address:\ ]{andre.xuereb@soton.ac.uk}
\author{Peter Domokos}
\affiliation{Research Institute of Solid State Physics and Optics, Hungarian Academy of Sciences, H-1525 Budapest P.O. Box 49, Hungary}
\author{Peter Horak}
\affiliation{Optoelectronic Research Centre, University of Southampton, Southampton SO17~1BJ, United Kingdom}
\author{Tim Freegarde}
\affiliation{School of Physics and Astronomy, University of Southampton, Southampton SO17~1BJ, United Kingdom}
\date{\today}

\begin{abstract}
We present a generic transfer matrix approach for the description of the interaction of atoms possessing multiple ground state and excited state sublevels with light fields. This model allows us to treat multi-level atoms as classical scatterers in light fields modified by, in principle, arbitrarily complex optical components such as mirrors, resonators, dispersive or dichroic elements, or filters. We verify our formalism for two prototypical sub-Doppler cooling mechanisms and show that it agrees with the standard literature.
\end{abstract}

\maketitle

\section{Introduction}\label{sec:Introduction}
The two-level model of atoms interacting with light fields~\cite{Shore1990a} has often been used to explore optical cooling mechanisms~\cite{Gordon1980,Dalibard1985a,Metcalf2003}. Its inherent simplicity---the atom has one ground state and one excited state---makes the resulting models amenable to analysis, but also suppresses mechanisms~\cite{Dalibard1989} that, in the appropriate parameter regimes, dominate the interaction.\par
A notable example of such an initially overlooked mechanism in atomic physics is three-dimensional optical molasses~\cite{Chu1985}. By means of the two-level model, one can predict the equilibrium temperature, the so-called ``Doppler'' temperature $T_D$, of atoms in molasses to be $\hbar\Gamma$, where $\Gamma$ is the (half-width at half-maximum) linewidth of the transition from the excited to the ground level~\cite{Metcalf2003}. Data from early three-dimensional molasses experiments contradicted this~\cite{Lett1988}, showing that the achievable equilibrium temperature was in fact much lower. This discrepancy was resolved independently by two groups~\cite{Dalibard1989,Ungar1989}, both explanations relying on the inclusion of the manifold of magnetic sublevels in each of the ground and excited states. In particular, the motion of the atoms in the optical field leads to a non-adiabatic following of the magnetic sublevel populations, which gives rise to a strong viscous force and efficient cooling to temperatures significantly lower than the Doppler temperature.
\par
We recently~\cite{Xuereb2009b} explored a new scattering theory that deals with the interaction of light and matter in a unified form applicable from microscopic to macroscopic systems. In that work we only considered the two-level atom model and showed, in particular, how our model can explain such mechanisms as standard optical molasses and mirror-mediated cooling~\cite{Xuereb2009a}. In this paper we extend this model to deal with magnetic sublevels, in much the same spirit as ref.~\cite{Dalibard1989}. In due course, this extension will enable us to deal with multilevel atoms interacting with an arbitrarily complex system composed of immobile mirrors, cavities, MEMS devices, etc., without resorting to a quantized model for such a system.
\\
After we introduce the general extension in the next section, we then proceed to explore two prototypical systems---the $J=\tfrac{1}{2}\rightarrow J^\prime=\tfrac{3}{2}$ transition, leading to the ``Sisyphus'' cooling mechanism, and the $J=1\rightarrow J^\prime=2$ transition---in \Sref{sec:LinPerpLin} and \Sref{sec:SigmaSigma}, respectively.

\section{A transfer matrix relating Jones vectors}\label{sec:Extension}
\begin{figure}[t]
 \centering
    \includegraphics[width=0.3\figwidth]{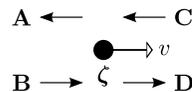}
\caption{Moving scatterer interacting with four field modes represented by the Jones vectors $\mathbf{A}$, $\mathbf{B}$, $\mathbf{C}$, and $\mathbf{D}$. The scatterwe has velocity $v$ and is described by means of its polarizability tensor $\boldsymbol{\zeta}$. The field mode amplitudes are, in general, functions of the wavenumber $k$.}
 \label{fig:System}
\end{figure}
We investigate the interaction of atoms with light of different polarizations. To this end, we denote the two polarization basis vectors by $\mu$ and $\nu$, whereby the standard circular polarization basis is equivalent to setting $\mu=\sigma^+$ and $\nu=\sigma^-$. Starting from the transfer matrix model explored in ref.~\cite{Xuereb2009b} and using the definitions in \fref{fig:System}, we replace each of the field modes by a corresponding Jones vector, similar to the model used in ref.~\cite{Spreeuw1992}. Thus, for example,
\begin{equation}
 A(k)\rightarrow\mathbf{A}(k)=\begin{pmatrix}
                               A_{\mu}(k)\\
                               A_{\nu}(k)
                              \end{pmatrix}\,,
\end{equation}
and similarly for $B$, $C$ and $D$. The transfer matrix $M$, describing the effect of the scatterer on the four field modes by means of the relation
\begin{equation}
\label{eq:StandardTMM}
 \begin{pmatrix}
    A(k)\\
    B(k)
  \end{pmatrix}=M
 \begin{pmatrix}
    C(k)\\
    D(k)
  \end{pmatrix}\,,
\end{equation}
is now transformed into an order $4$ tensor of the form
\begin{equation}
 \boldsymbol{M}=\begin{bmatrix}
    \boldsymbol{m_{11}} & \boldsymbol{m_{12}}\\
    \boldsymbol{m_{21}} & \boldsymbol{m_{22}}
   \end{bmatrix}\,,
\end{equation}
where each of $\boldsymbol{m_{\alpha\beta}}$ ($\alpha,\beta=1,2$) is a $2\times 2$ matrix relating the respective Jones vector components. A general recipe for transforming the formulae for the field mode amplitudes, as given in ref.~\cite{Xuereb2009b}, can be summarized by means of the two replacements
\begin{equation}
 1\rightarrow\boldsymbol{1}=\begin{bmatrix}
                 1 & 0\\
                 0 & 1
                \end{bmatrix}\text{ and }\zeta\rightarrow\boldsymbol{\zeta}\,,
\end{equation}
wherever necessary. In particular, then,
\begin{equation}
\label{eq:NewMatrix}
 M=\begin{bmatrix}
 1-i\zeta & -i\zeta\\
 i\zeta & 1+i\zeta
\end{bmatrix}\rightarrow\boldsymbol{M}=\begin{bmatrix}
 \boldsymbol{1}-i\boldsymbol{\zeta} & -i\boldsymbol{\zeta}\\
 i\boldsymbol{\zeta} & \boldsymbol{1}+i\boldsymbol{\zeta}
\end{bmatrix}\,.
\end{equation}
We follow ref.~\cite{CohenTannoudji1978a}, Complement E$_\textrm{III}$ \textsection 3-b, in defining the polarizability tensor $\boldsymbol{\zeta}$ as the steady-state expectation value of the polarizability operator $\hat{\boldsymbol{\chi}}$; $\boldsymbol{\zeta}$ is therefore given by the trace
\begin{equation}
 \boldsymbol{\zeta}=\Tr\bigl(\rho^\text{st}\cdot\hat{\boldsymbol{\chi}}\bigr)=\sum_{i,j}\langle j\rvert\rho^\text{st}\lvert i\rangle\langle i\rvert\hat{\boldsymbol{\chi}}\lvert j\rangle\,,
\end{equation}
where $\rho^\text{st}$ is the steady-state density matrix describing the system and the summation runs over all the internal sublevels of the atom, and where we construct the order 4 polarizability operator tensor $\hat{\boldsymbol{\chi}}$ similarly to ref.~\cite{Shore1990b}, Eq.~(14.9-24). In the general $\mu$, $\nu$ basis:
\begin{align}
 \label{eq:ChiME}
 \langle i\rvert\hat{\boldsymbol{\chi}}\lvert j\rangle
                       &=\zeta_0\sum_e\Biggl(\begin{matrix}
                                    \langle i\rvert\hat{d}_\mu\lvert e\rangle\\
                                    \langle i\rvert\hat{d}_\nu\lvert e\rangle
                                   \end{matrix}\Biggr)\otimes
                                   \Biggl(\begin{matrix}
                                    \langle j\rvert\hat{d}_\mu\lvert e\rangle\\
                                    \langle j\rvert\hat{d}_\nu\lvert e\rangle
                                   \end{matrix}\Biggr)\nonumber\\
                       &=\zeta_0\sum_e\Biggl[\begin{matrix}
                                    \langle i\rvert\hat{d}_\mu\lvert e\rangle\langle e\rvert\hat{d}_\mu\lvert j\rangle & \langle i\rvert\hat{d}_\mu\lvert e\rangle\langle e\rvert\hat{d}_\nu\lvert j\rangle\\
                                    \langle i\rvert\hat{d}_\nu\lvert e\rangle\langle e\rvert\hat{d}_\mu\lvert j\rangle & \langle i\rvert\hat{d}_\nu\lvert e\rangle\langle e\rvert\hat{d}_\nu\lvert j\rangle
                                   \end{matrix}\Biggr]\,,
\end{align}
with $\zeta_0$ being the characteristic polarizability of the atom. In \eref{eq:ChiME}, the dipole moment operator $\hat{d}_\mu$ ($\hat{d}_\nu$) is related to the $\mu$ ($\nu$) polarized light field and the sum runs over all the internal sublevels, $e$, of the atom. The matrix elements of $\hat{d}_\mu$ ($\hat{d}_\nu$) are given by the appropriate Clebsch-Gordan coefficients.
\par
Importantly, this new transfer matrix still retains all its properties, allowing us to model the interaction of the multilevel atom with an arbitrary system of immobile optical elements such as mirrors, cavities, waveplates, etc. As in our previous work~\cite{Xuereb2009b}, this interaction is accounted for by the multiplication of the various transfer matrices of the elements making up the system; this model is, in principle, applicable to systems of arbitrary complexity.
\par
Finally, we recall that the diagonal elements, $\langle i\rvert\rho^\text{st}\lvert i\rangle$, of $\rho^\text{st}$ are the populations in each of the sublevels, whereas its off-diagonal elements, $\langle i\rvert\rho^\text{st}\lvert j\rangle$, are the respective coherences. The matrix elements of $\rho^\text{st}$ are obtained from the appropriate optical Bloch equations (see, for example, the procedure outlined in ref.~\cite{CohenTannoudji1977b}). We note here that, through its dependence on $\rho^\text{st}$, $\boldsymbol{M}$ depends on the fields that it helps to determine, and thus \eref{eq:StandardTMM} will in general become a set of nonlinear equations. In cases, like the ones considered in the following sections, where only one multilevel atom is interacting with a linear optical system, this problem may be solved using a procedure similar to the one outlined below: the fields surrounding the atom are obtained from the input fields through linear operations and then used with the optical Bloch equations to obtain the populations and coherences of the atom's various levels. Knowledge of these quantities then determines the fields, and hence the forces acting on the atom, completely.
\ \\\par
In the following sections we will restrict our discussion to the case where the input field is not modified by other transfer matrices. We will apply this mechanism to investigate the behaviour of atoms in two cases where the polarization of the light varies in space on scales of the order of the wavelength to verify the validity of the model given by \eref{eq:NewMatrix} to \eref{eq:ChiME}. In the first instance, we illuminate our atom with two counterpropagating linearly polarized beams. We choose the planes of polarization of the two beams to be orthogonal to each other. The second configuration we will investigate involves illuminating the atom with two circularly polarized beams, choosing opposite handedness for the two beams. These two cases mirror those in ref.~\cite{Dalibard1989}.

\section{Atoms in a gradient of polarization}\label{sec:LinPerpLin}
\begin{figure}[t]
 \centering
    \includegraphics[scale=0.55]{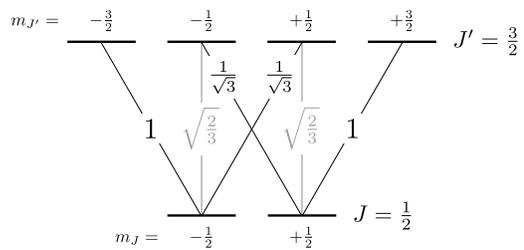}
\caption{Clebsch-Gordan coefficients for a $J=\tfrac{1}{2}\rightarrow J^\prime=\tfrac{3}{2}$ transition.}
 \label{fig:1-2_to_3-2}
\end{figure}
In this and the following sections, we will adopt the low-intensity hypothesis. This allows us to simplify the optical Bloch equations and resulting system considerably by neglecting the populations and coherences of the excited state sublevels. We can thus replace $\rho^\text{st}$ by the ground state steady-state density matrix, $\rho_\text{g}^\text{st}$. We denote the diagonal element $(i,i)$ of $\rho_\text{g}^\text{st}$, the population in sublevel $i$, by $\Pi_i$, and the off-diagonal element $(i,j)$, the coherence between sublevels $i$ and $j$, by $C_{i,j}$.
\par
Here we will discuss what is perhaps the simplest transition between two levels with multiple magnetic sublevels: the $J=\tfrac{1}{2}\rightarrow J^\prime=\tfrac{3}{2}$ transition. In this case, we have two ground sublevels so that $\rho_g^\text{st}$ is a $2\times 2$ matrix. \Fref{fig:1-2_to_3-2} tabulates the Clebsch-Gordan coefficients required to evaluate $\boldsymbol{\zeta}$. We thus have:
\begin{equation}
 \rho_g^\text{st} = \begin{bmatrix}
                     \Pi_{-\tfrac{1}{2}}             & C_{-\tfrac{1}{2},+\tfrac{1}{2}}\\
                     C_{+\tfrac{1}{2},-\tfrac{1}{2}} & \Pi_{+\tfrac{1}{2}}
                    \end{bmatrix}
\end{equation}
and
\begin{equation}
 \hat{\boldsymbol{\chi}}=\zeta_0\begin{pmatrix}
                 \begin{bmatrix}
                     \tfrac{1}{3} & 0\\
                     0 & 1
                 \end{bmatrix} &
                 \boldsymbol{0}\\
                 \boldsymbol{0}&
                 \begin{bmatrix}
                     1 & 0\\
                     0 & \tfrac{1}{3}
                 \end{bmatrix}
                \end{pmatrix}\,,
\end{equation}
whereby
\begin{equation}
 \boldsymbol{\zeta}=\zeta_0\Biggl(\begin{bmatrix}
                     \tfrac{1}{3} & 0\\
                     0 & 1
                 \end{bmatrix}\Pi_{-\tfrac{1}{2}}+
                 \begin{bmatrix}
                     1 & 0\\
                     0 & \tfrac{1}{3}
                 \end{bmatrix}\Pi_{+\tfrac{1}{2}}\Biggr)\,.
\end{equation}
Suppose, now, that we illuminate the atom with two counterpropagating beams having orthogonal linear polarization and equal intensity. This can be represented by setting
\begin{equation}
 \mathbf{B}(k)=\frac{B}{\sqrt{2}}\begin{pmatrix}
                     1\\
                     1
                   \end{pmatrix}\,\exp(ikx-i\pi/4)\end{equation}
and
\begin{equation}
 \mathbf{C}(k)=\frac{iB}{\sqrt{2}}\begin{pmatrix}
                     1\\
                     -1
                   \end{pmatrix}\exp(-ikx+i\pi/4)\,,
\end{equation}
where the shift in the $x$ coordinate is introduced to simplify our expressions. Using the optical Bloch equations, we can show that the steady state populations in the ground sublevels at zero atomic velocity are given by
\begin{equation}
\Pi_{-\tfrac{1}{2}}=\cos^2(kx)\,\text{ and }\,\Pi_{+\tfrac{1}{2}}=\sin^2(kx)\,,
\end{equation}
noting that the populations do not depend on the field amplitudes in the low intensity regime.
\par
We work to lowest order in $\zeta_0$ and make use of the above relations to find the net force acting on the atom:
\begin{align}
\label{eq:GeneralForce}
 \force&=\hbar k\Bigl(\lvert\mathbf{A}\rvert^2+\lvert\mathbf{B}\rvert^2-\lvert\mathbf{C}\rvert^2-\lvert\mathbf{D}\rvert^2\Bigr)\nonumber\\
&=2\hbar k\im{\bigl[\boldsymbol{\zeta}\bigl(\mathbf{B}+\mathbf{C}\bigr)\bigr]\cdot\bigl(\mathbf{B}-\mathbf{C}\bigr)^\star}\nonumber\\
&\phantom{=\ }+4\tfrac{v}{c}\hbar k\im{\bigl(\boldsymbol{\zeta}\mathbf{B}\bigr)\cdot\mathbf{C}^\star+\bigl(\boldsymbol{\zeta}\mathbf{C}\bigr)\cdot\mathbf{B}^\star}\nonumber\\
&\phantom{=\ }-2\tfrac{v}{c}\hbar k^2\im{\biggl[\frac{\partial\boldsymbol{\zeta}}{\partial k}\bigl(\mathbf{B}+\mathbf{C}\bigr)\biggr]\cdot\bigl(\mathbf{B}+\mathbf{C}\bigr)^\star}\nonumber\\
&\approx2\hbar k\lvert B\rvert^2\im{\bigl[\boldsymbol{\zeta}\bigl(\mathbf{B}+\mathbf{C}\bigr)\bigr]\cdot\bigl(\mathbf{B}-\mathbf{C}\bigr)^\star}\nonumber\\
&\phantom{\approx\ }+2\tfrac{v}{c}\hbar k^2\im{\biggl[\frac{\partial\boldsymbol{\zeta}}{\partial k}\bigl(\mathbf{B}+\mathbf{C}\bigr)\biggr]\cdot\bigl(\mathbf{B}+\mathbf{C}\bigr)^\star}\,,
\end{align}
where we have assumed that $\lVert k\,\partial\boldsymbol{\zeta}/\partial k\rVert\gg \lVert\boldsymbol{\zeta}\rVert$. The velocity-dependent force terms in the above expression arise through the Doppler shifting of photons both between field modes in the same polarization and between field modes in different polarizations; these mechanisms are accounted for by the diagonal and off-diagonal terms in $\boldsymbol{\zeta}$, respectively. These terms emerge through the velocity-dependent terms in the generalised transfer matrix.
\\
In the present case, \eref{eq:GeneralForce} simplifies approximately to
\begin{align}
 \force&=\tfrac{4}{3}\hbar k\zeta_0\lvert B\rvert^2\sin(2kx)\Bigl(\Pi_{+\tfrac{1}{2}}-\Pi_{-\tfrac{1}{2}}\Bigr)\nonumber\,,
\end{align}
assuming that $\zeta_0$ is real for simplicity.
\\
We now let $\tau_p$ be a characteristic residence time of the two ground state sublevels; this will introduce a non-adiabatic following term, proportional to $v$, in the populations of each of the sublevels and emerges from the optical Bloch equations. Thus, we obtain the expression
\begin{equation}
 \force=-\tfrac{2}{3}\hbar k\lvert B\rvert^2\zeta_0\sin(4kx)-\tfrac{8}{3}\hbar k^2\lvert B\rvert^2\zeta_0v\tau_\text{p}\sin^2(2kx),
\end{equation}
which agrees precisely with the standard literature (cf. Eqs.~(4.20) and~(4.23) in ref.~\cite{Dalibard1989}).

\section{Atoms in a gradient of ellipticity}\label{sec:SigmaSigma}
\begin{figure}[t]
 \centering
    \includegraphics[scale=0.55]{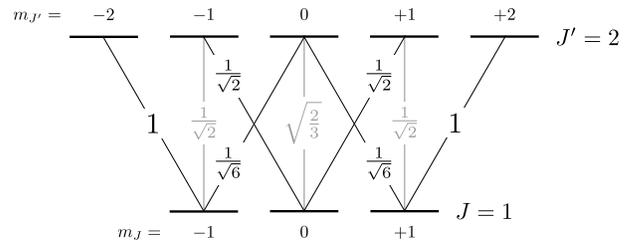}
\caption{Clebsch-Gordan coefficients for a $J=1\rightarrow J^\prime=2$ transition.}
 \label{fig:1_to_2}
\end{figure}
If we illuminate an atom with two counterpropagating beams of light in a $\sigma^+$--$\sigma^-$ configuration, rich dynamics are obtained not in the simplest ($J=\tfrac{1}{2}\rightarrow J^\prime=\tfrac{3}{2}$) case, but in the next simplest, where the ground state has three magnetic sublevels ($J=1$) and the excited state five ($J^\prime=2$). In this case, then, we can express $\rho_g^\text{st}$ and $\hat{\boldsymbol{\chi}}$ as
\begin{equation}
 \rho_g^\text{st} = \begin{bmatrix}
                     \Pi_{-1}  & C_{-1,0} & C_{-1,+1}\\
                     C_{0,-1}  & \Pi_0    & C_{0,+1}\\
                     C_{+1,-1} & C_{+1,0} & \Pi_{+1}
                    \end{bmatrix}
\end{equation}
and
\begin{equation}
 \hat{\boldsymbol{\chi}}=\zeta_0\begin{pmatrix}
                 \begin{bmatrix}
                     \tfrac{1}{6} & 0\\
                     0            & 1
                 \end{bmatrix} &
                 \boldsymbol{0} &
                 \begin{bmatrix}
                     0 & \tfrac{1}{6}\\
                     0 & 0
                 \end{bmatrix}\\
                 \boldsymbol{0} &
                 \begin{bmatrix}
                     \tfrac{1}{2} & 0\\
                     0            & \tfrac{1}{2}
                 \end{bmatrix} &
                 \boldsymbol{0}\\
                 \begin{bmatrix}
                     0            & 0\\
                     \tfrac{1}{6} & 0
                 \end{bmatrix}&
                 \boldsymbol{0} &
                 \begin{bmatrix}
                     1 & 0\\
                     0 & \tfrac{1}{6}
                 \end{bmatrix}
                \end{pmatrix}\,,
\end{equation}
using the Clebsch-Gordan coefficients in \fref{fig:1_to_2}. Together, these give
\begin{align}
 \boldsymbol{\zeta}=\zeta_0\Biggl(&\begin{bmatrix}
                     \tfrac{1}{6} & 0\\
                     0            & 1
                 \end{bmatrix}\Pi_{-1}+
                 \begin{bmatrix}
                     \tfrac{1}{2} & 0\\
                     0            & \tfrac{1}{2}
                 \end{bmatrix}\Pi_0+
                 \begin{bmatrix}
                     1 & 0\\
                     0 & \tfrac{1}{6}
                 \end{bmatrix}\Pi_{+1}\nonumber\\
                 &+\begin{bmatrix}
                     0 & \tfrac{1}{6}\\
                     0 & 0
                 \end{bmatrix}C+
                 \begin{bmatrix}
                     0            & 0\\
                     \tfrac{1}{6} & 0
                 \end{bmatrix}C^\star\Biggr)\,,
\end{align}
 with $C=C_{+1,-1}=C_{-1,+1}^\star=\langle+1\vert\rho_\text{g}^\text{st}\vert\!-\!\!1\rangle$ representing the nonzero coherence between the $m_J=+1$ and the $m_J=-1$ sublevels. Note that we again apply the low intensity hypothesis, thereby replacing $\rho^\text{st}$ with $\rho_\text{g}^\text{st}$.
\par
We now illuminate the atom with two counterpropagating beams of equal intensity, $\mathbf{B}$ and $\mathbf{C}$, possessing $\sigma^+$ and $\sigma^-$ polarization, respectively:
\begin{equation}
 \mathbf{B}(k)=B\begin{pmatrix}
                     1\\
                     0
                   \end{pmatrix}\,\exp(ikx)
\end{equation}
and
\begin{equation}
 \mathbf{C}(k)=B\begin{pmatrix}
                     0\\
                     1
                   \end{pmatrix}\exp(-ikx)\,.
\end{equation}
We again use~\eref{eq:GeneralForce} to derive the force acting on the atom, which is given by
\begin{align}
\label{eq:SigmaSigmaForce}
 \force&=2\hbar k\lvert B\rvert^2\,\mathrm{Im}\bigl\{\tfrac{5}{6}\zeta_0\bigl(\Pi_{+1}-\Pi_{-1}\bigr)\nonumber\\
&\phantom{=2\hbar k\lvert B\rvert^2\,\mathrm{Im}\bigl\{}+\tfrac{1}{6}i\zeta_0\im{C\exp(-2ikx)}\bigr\}\nonumber\\
&\phantom{=\ }-2\tfrac{v}{c}\hbar k^2\lvert B\rvert^2\im{\partial\zeta_0/\partial k}\Bigl(\tfrac{7}{6}\bigl(\Pi_{+1}+\Pi_{-1}\bigr)+\Pi_0\nonumber\\
&\phantom{=\ -2\tfrac{v}{c}\hbar k^2\lvert B\rvert^2\im{\partial\zeta_0/\partial k}\Bigl(}+\tfrac{1}{3}\re{C\exp(-2ikx)}\Bigr)\,,
\end{align}
where the populations and coherences are again obtained from the optical Bloch equations, and can be found in ref.~\cite{Dalibard1989}. By observing the natural correspondence between $\zeta_0$ and $s_\pm$ in this latter reference, we can see that our expression for the force acting on the atom again agrees with the standard literature to first order in $\tfrac{v}{c}$ (cf. Eq.~(5.9) in ref.~\cite{Dalibard1989}).
The resulting friction force is thus due to both the Doppler shift, as evident in the terms shown explicitly in~\eref{eq:SigmaSigmaForce}, as well as through the non-adiabatic following of the atomic sublevel populations.

\section{Conclusions}\label{sec:Conclusions}
By revisiting the transfer matrix formalism and expressing the polarizability of a scatterer as the expectation value of a quantum operator, we have endowed it with a strong quantum character that allows us to handle atoms with multiple ground and excited state sublevels. In principle, our extended formalism is only limited by its reliance on the optical Bloch equations to give expressions for the ground state populations and coherences; we have retained the character of our earlier formalism that allowed us to work to arbitrary order in the polarizability. We have applied this theory to two standard sub-Doppler cooling configurations, the so-called ``lin--$\perp$--lin'' and ``$\sigma^+$--$\sigma^-$'' configurations, and thereby reproduced the known expressions for the force acting on the atom.

\section*{Acknowledgements}
This work was supported by the UK Engineering and Physical Sciences Research Council (EPSRC) grant EP/E058949/1 and by the \emph{Cavity-Mediated Molecular Cooling} network within the EuroQUAM programme of the European Science Foundation (ESF), as well as by the National Scientific Fund of Hungary (Contract No. NF68736).

\end{document}